\documentclass[aps,prd,reprint,preprintnumbers,superscriptaddress,nofootinbib,longbibliography,floatfix]{revtex4-2}
\pdfoutput=1

\usepackage[utf8]{inputenc} 
\usepackage[T1]{fontenc}    
\usepackage{hyperref}       
\usepackage{url}            
\usepackage{booktabs}       
\usepackage{amsfonts}       
\usepackage{nicefrac}       
\usepackage{microtype}      
\usepackage{xcolor}         
\usepackage{amsmath}
\usepackage{graphicx}
\usepackage{float}
\usepackage{subfig}
\usepackage{amsmath}
\usepackage{physics}
\usepackage{xcolor}
\usepackage{multirow}

\hypersetup{
  colorlinks=true,
  citecolor=blue,
  linkcolor=blue,
  urlcolor=blue
}

\usepackage{physics-defs}
\usepackage{placeins}

\renewcommand{\eta}{y}
\usepackage{physics-defs}

\begin{document}

\title{Parton Labeling without Matching: \\ Unveiling Emergent Labelling Capabilities in Regression Models}

\author{Shikai Qiu}
\email{calvin_qiu@berkeley.edu}
\affiliation{Department of Physics, University of California, Berkeley, Berkeley, CA 94720, USA}
\affiliation{Courant Institute of Mathematical Sciences, New York University, New York, NY 10012}

\author{Shuo Han}
\email{shuohan@lbl.gov}
\affiliation{Physics Division, Lawrence Berkeley National Laboratory, Berkeley, CA 94720, USA}

\author{Xiangyang Ju}
\email{xju@lbl.gov}
\affiliation{Physics Division, Lawrence Berkeley National Laboratory, Berkeley, CA 94720, USA}

\author{Benjamin Nachman}
\email{bpnachman@lbl.gov}
\affiliation{Physics Division, Lawrence Berkeley National Laboratory, Berkeley, CA 94720, USA}
\affiliation{Berkeley Institute for Data Science, University of California, Berkeley, CA 94720, USA}

\author{Haichen Wang}
\email{haichenwang@berkeley.edu}
\affiliation{Physics Division, Lawrence Berkeley National Laboratory, Berkeley, CA 94720, USA}
\affiliation{Department of Physics, University of California, Berkeley, Berkeley, CA 94720, USA}

\begin{abstract}

Parton labeling methods are widely used when reconstructing collider events with top quarks or other massive particles.  State-of-the-art techniques are based on machine learning and require training data with events that have been matched using simulations with truth information.  In nature, there is no unique matching between partons and final state objects due to the properties of the strong force and due to acceptance effects.  We propose a new approach to parton labeling that circumvents these challenges by recycling regression models.  The final state objects that are most relevant for a regression model to predict the properties of a particular top quark are assigned to said parent particle without having any parton-matched training data.  This approach is demonstrated using simulated events with top quarks and outperforms the widely-used $\chi^2$ method.

\end{abstract}

\maketitle

\clearpage

\clearpage

\section{Introduction}

A common task in collider event reconstruction is assigning final state objects to a branch of the hypothesized reaction that generated the event.  For example, hard-scatter events with outgoing quarks and gluons produce jets that can be associated with their initiating partons.  When there are many outgoing particles from the hard-scatter reaction, this is a complex combinatorial challenge.  Events with multiple top quarks naturally result in such final states, since nearly all top quarks decay to a $b$-quark and a $W$ boson, which subsequently decays to two quarks or leptons.  A key challenge in many measurements and searches involving top quarks is the assignment of reconstructed objects with one of the top quark decay products.  Classically, this assignment has used $\chi^2$ or related methods that enumerate all possibilities and pick the one which is most consistent with having two on-shell $W$ boson and top quark intermediaries. The difficulty with these methods is that they do not take into account all available information and are computationally expensive. 

A number of modern machine learning (ML) methods have been proposed to address these challenges.  These techniques range from Boosted Decision Trees~\cite{ATLAS:2017fak,CMS:2019eih,ATLAS:2020ior} and existing neural networks~\cite{Erdmann:2019evj,Badea:2022dzb} to custom, permutation invariant deep learning methods~\cite{Fenton:2020woz, Lee:2020qil, Shmakov:2021qdz,Ehrke:2023cpn}. In all cases, object identification can make use of a variety of lepton-, jet- and event-level properties that were inaccessible with $\chi^2$ or likelihood methods~\cite{Erdmann:2013rxa}.  This is possible because the ML approaches are trained on simulations, so whatever information is available and well-modeled (within uncertainty) can be used for object labeling.

\begin{figure}
    \centering
    \includegraphics[width=0.48\textwidth]{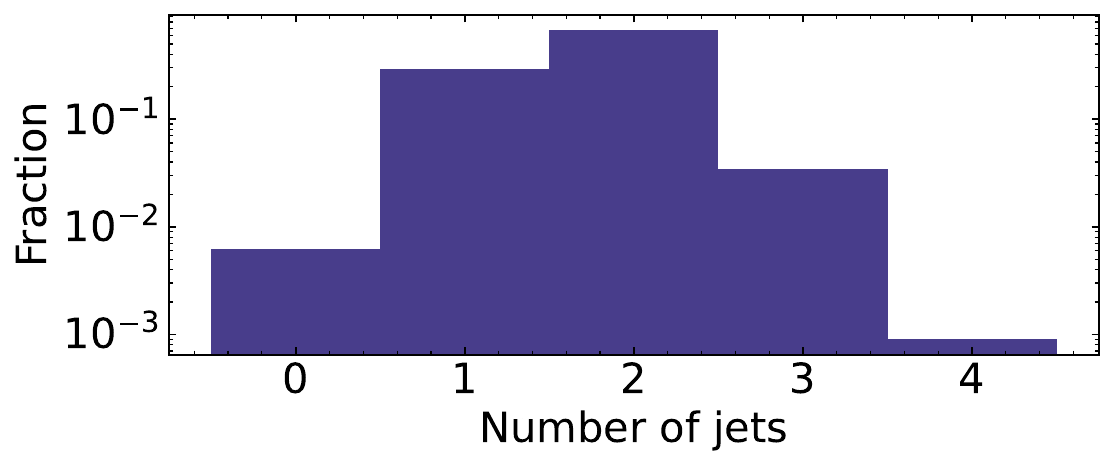}\\
    \includegraphics[width=0.48\textwidth]{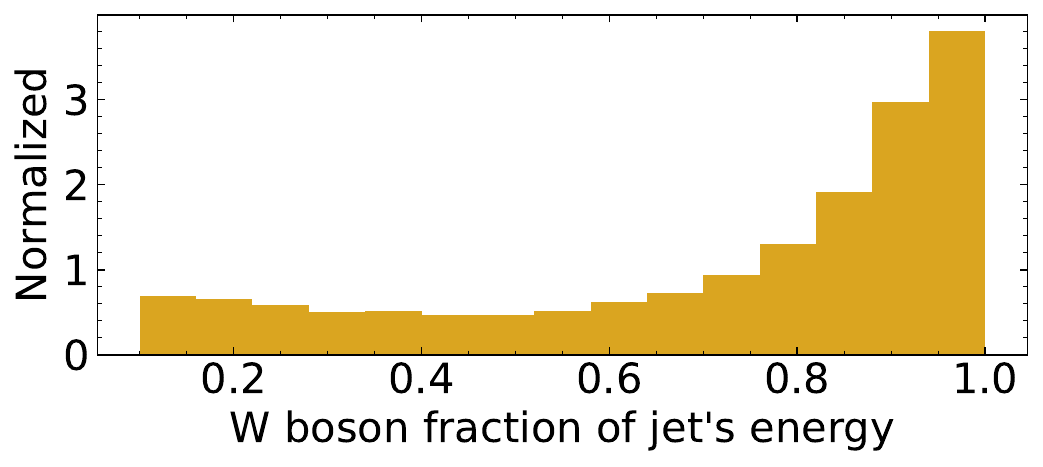}
    \caption{Simulated all-hadronic $t\bar{t}$ events.  In the $N_\text{colors}\rightarrow\infty$ limit, hadrons can be uniquely associated as $W$ boson descendants.  Top: number of jets with at least 10\% of their energy from the $W^+$.  Bottom: of these jets, the fraction of their energy from the $W^+$.   Jets are clustered using the anti-$k_t$~\cite{Cacciari:2008gp} algorithm with $R=0.4$.}
    \label{fig:wjets}
\end{figure}

Despite the success of these ML methods, they all share a common fundamental challenge with classical approaches.  In particular, they all require matched objects for training.  This may be problematic for two reasons (see e.g. Fig.~\ref{fig:wjets}).  First, there is no unique match between a hard-scatter quark/gluon and a jet.  A single quark/gluon can fragment into multiple jets, and a single jet can be composed of hadrons with energy flow originating from multiple quarks/gluons.  This is particularly acute for top quarks, which carry color charge and thus must be color-connected to another quark/gluon in the event.  The extent of the overlap also depends on the jet clustering algorithm - jets with a larger catchment area~\cite{Cacciari:2008gn} are more likely to be due to the merger of multiple parton showers.  Second, even if a parent object like a top quark could be uniquely associated with a set of decay products, acceptance effects will obscure the association.  In particular, the finite geometric and energy acceptance of detectors results in missed final state objects.

Our philosophy is to circumvent the issues caused by object-parton matching by directly regressing onto the target particle properties.  In Ref.~\cite{Qiu:2022xvr}, we designed the Covariant Particle Transformer (CPT), a partially Lorentz covariant point cloud transformer, to learn the four-vectors of top quarks given reconstructed jets, leptons, photons, and missing energy.  In this paper, we show how one can reuse such a regression method to perform parton labeling.  We explore two possibilities, one based on the attention mechanism within the CPT and one based on the gradient of predicted four-vectors with respect to the inputs.  The latter approach is compatible with any regression-based top quark reconstruction method, even if it does not involve neural network attention.  While we still advocate for regression in cases where the underlying top quark properties are needed, parton labeling is still widely used for determining these properties and no matter what approach is used, parton labels can be useful for diagnostic purposes.

This paper is organized as follows.  Section~\ref{sec:methods} briefly reviews the CPT technique and then introduces our two approaches to extracting parton labels from the regression model.  Numerical results are presented in Sec.~\ref{sec:results} using a dataset that is briefly introduced in Sec.~\ref{sec:data}.  The paper ends with conclusions and outlook in Sec.~\ref{sec:end}.

\section{Methods}
\label{sec:methods}

Our goal is to take final states with $n$ top quarks that decay hadronically and assign jets to one of these quarks.  In principle, one could simultaneously predict $n$ and assign jets, but in practice, there is often a particular number of target quarks; if not, one could first run a multi-class classification procedure.  We also restrict our approach to assigning three jets to each top quark. Both ML-based approaches described below could be modified to assign fewer or more jets by placing thresholds on the Jacobean values (Sec.~\ref{sec:grad}) or the attention weights (Sec.~\ref{sec:att}), but we leave this to future work. 

\subsection{Covariant Particle Transformer}
\label{sec:CPT}
The Covariant Particle Transformer (CPT) is a Transformer-based \cite{vaswani2017attention} neural network tailored for collider physics applications and has demonstrated superior performance in predicting top quarks' kinematics compared to classical approaches \cite{Qiu:2022xvr}. CPT takes as inputs the 4-vectors and particle identifications of all observed final state objects (jets, lepton, photons, etc.) and outputs predicted 4-vectors of a pre-specified number of top quarks. Compared to the standard transformer architecture, CPT is designed to respect important symmetries in collider physics: it is permutation invariant under reordering of the inputs and partially Lorentz covariant, meaning if we apply a longitudinal boost and/or a transverse rotation to all the inputs, CPT's outputs will be boosted and/or rotated accordingly, respecting Lorentz symmetry. 

In each layer of the network, CPT additively updates the feature vector of every object $f_i$ (could be an input or output) with $\Delta f_i$ defined as a function of all the feature vectors $\{f_k\}:$
\begin{equation}
    \label{eq:attention}
    \Delta f_i = \sum_{k} \alpha_{ik} \varphi(f_k),
\end{equation}
where $\varphi$ is a learned linear transformation and $\{\alpha_{ik}\}$ are positive attention weights, which are themselves non-linear functions of $\{f_k\},$ such that $\sum_k \alpha_{ik} = 1$ for each $i.$ The output feature vectors are eventually transformed to the predicted 4-vectors of the top quarks. If $i$ is an output index and $k$ is an input index, then intuitively $\alpha_{ik}$ measures the importance of the information in $k$ for predicting the properties of $i.$ The above procedure is named the covariant attention mechanism, which modifies the standard attention mechanism in a transformer to ensure partial Lorentz covariance. To capture complex correlations between the inputs and outputs,
CPT uses $L=6$ covariant attention layers and $H=4$ attention heads per layer to decode the top quark 4-vectors, where each attention head performs separate learned updates according to Equation~\ref{eq:attention} for added flexibility. We refer readers to the original CPT paper for a more comprehensive review of the architecture and implementation. 

\subsection{Gradient-based Labeling}
\label{sec:grad}

The idea of the gradient-based method is to assign a jet to a particular top quark if changes to the jet properties result in significant changes to the top quark properties.  If the top quarks were produced independently of each other and of other radiation within the event, then only the jets they produce should be relevant for reconstructing their properties.  In reality, this is not the case because top quarks and other objects are correlated through momentum conservation and other physics effects.

Strictly speaking, the term `gradient' applies to the case of one-dimensional quantities (e.g. top quark $p_T$), but for regression methods that predict multiple top quark properties, a more accurate name would be `Jacobian-based'.  For simplicity, we will henceforth always call this method `gradient-based'.

The gradient-based labeling scheme is compatible with any regression model (not just the CPT from Sec.~\ref{sec:CPT}) and is based on the following quantity:

\begin{align}
\label{eq:grad}
    \Delta_{ik}=\norm{\left(\frac{\partial f_{i,p_T}}{\partial j_{k,p_T}},\frac{\partial f_{i,\eta}}{\partial j_{k,\eta}},\frac{\partial f_{i,\phi}}{\partial j_{k,\phi}}\right)},
\end{align}
where $f_{i,x}$ is the predicted $x\in\{p_T,\eta,\phi\}$ of top quark $i$ and $j_{k,x}$ is the observed $x$ of jet $k$.  Since $f_i$ is a neural network, we can compute the derivatives in Eq.~\ref{eq:grad} using the same automatic differentiation (e.g. back propagation) that is used when training the network in the first place.  We assign jet $k$ to top quark $i$ if $\Delta_{ik}$ is one of the top three values across all $k$.  The same jet could be assigned to multiple top quarks.  Equation~\ref{eq:grad} is not the unique combination of elements from the Jacobian and it could be that other combinations could be more effective.  We found that using the derivatives with respect to $p_T$, $y$, and $\phi$ was only slightly better than $p_T$ alone. More complex schemes that weight the different entries separately are also possible.

When $f$ is a CPT, then $\Delta_{ik}$ is a partial Lorentz scalar and so the labeling is invariant under longitudinal boosts and rotations in the transverse plane.

\subsection{Attention-based Labeling}
\label{sec:att}
In each covariant attention layer and attention head in CPT, the attention weight $\alpha_{ik}$ can be interpreted as a measure of the importance of input $k$ for predicting the properties of top $i,$ locally in the network. By averaging $\alpha_{ik}$ over all layers and attention heads, we obtain a measure of the overall importance of input $k$ to top $i$:
\begin{equation}
    \bar{\alpha}_{ik} = \frac{1}{LH}\sum_{\ell, h} \alpha^{\ell h}_{ik},
\end{equation}
where $\alpha^{\ell h}_{ik}$ is the attention weight between top $i$ and input $k$ in the $h^\text{th}$ attention head in the $\ell^\text{th}$ layer. Similar to gradient-based labeling, we assign the jet with index $k$ to top quark $i$ if $\bar{\alpha}_{ik}$ is one of the top three values across all jets. 

Due to the design of CPT, all attention weights are partial Lorentz scalars and $\bar{\alpha}_{ik}$ is again a partial Lorentz scalar, implying the labeling is invariant under longitudinal boosts and rotations in the transverse plane.

\subsection{$\chi^2$-based Labeling}

The baseline parton labeling scheme that we use is a widely applied $\chi^2$ method.  In particular, in events with at least two jets tagged as originating from bottom quarks ($b$-jets), the assignment of jets to top quarks is based on the combination that minimized the following $\chi^2$:

\begin{align}\nonumber
    \chi^2 &= \frac{(m_{b_1j_1j_2}-m_t)^2}{\sigma_{m_{bjj}}^2}+\frac{(m_{b_2j_3j_4}-m_t)^2}{\sigma_{m_{bjj}}^2}\\
   & \hspace{8mm}+\frac{(m_{j_1j_2}-m_W)^2}{\sigma_{m_{jj}}^2}+\frac{(m_{j_3j_4}-m_W)^2}{\sigma_{m_{jj}}^2}\,,
\end{align}
where $m_t$ and $m_W$ are the top quark and $W$ boson masses, respectively, and $\sigma_{m_{bjj}}$ and $\sigma_{m_{jj}}$ are the resolutions of truth-matched top and $W$ events, respectively.  As in this case, when we need to refer to classical truth labels, we will call top quarks that have all three decay products as `truth-matched' when each of the three quark decay products is within $\Delta R<0.4$ of exactly one jet (about 20\% efficient). Events without six jets, two of which are $b$-tagged, are not reconstructable with the $\chi^2$ method.  It may be possible to recover some of the non-reconstructable cases using other approaches for the $b$-jets (e.g. taking the highest energy jet(s)), so we check that our results hold in cases where events have two $b$-jets.

\section{Dataset}
\label{sec:data}

For numerical studies, we use the same dataset as in Ref.~\cite{Qiu:2022xvr}, which is briefly summarized below.  Top quark pair production in association with a Higgs boson\footnote{The Higgs boson decays to photons and is largely ignored and irrelevant for jet labeling.  We use this sample because it was the main one used in Ref.~\cite{Qiu:2022xvr}, although it was also shown that the performance is similar in other top quark final states.} in proton-proton collisions is generated with Madgraph@NLO~2.3.7~\cite{Alwall:2014hca} at next-to-leading order (NLO) in Quantum Chromodynamics (QCD). The decays of the top quarks are simulated with MadSpin~\cite{Artoisenet:2012st} and then the rest of the particle-level generation is created with Pythia 8.235~\cite{Sjostrand:2014zea}.   While this dataset does not emulate detector effects, the salient features of the problem are already present at particle level. Jets are clustered using the anti-$k_t$~\cite{Cacciari:2008gp} algorithm with $R=0.4$ as implemented in \textsc{FastJet} 3.3.2~\cite{Cacciari:2011ma,Cacciari:2005hq}. 

Jets are required to have $|\eta| \leq 2.5$ and $p_T \geq 25$~GeV.  Jets that are $\Delta R$ matched\footnote{$\Delta R$ is defined as $\sqrt{\Delta\eta ^2 + \Delta\phi^2}$, where $\Delta\eta$ is the difference of two particles in pseudorapidity and $\Delta\phi$ is the difference in azimuthal angle.} to $b$-quarks at the parton level are labeled as $b$-jets; this label is removed\footnote{We do not add fake $b$-jets, since the fake rate (one in a few hundred) is sufficiently small that missing a real $b$-jet and falsely tagging a non $b$-jets is rare enough to not impact the numerical results.} randomly for 30\% of the $b$-jets, to mimic the inefficiency of a realistic $b$-tagging~\cite{ATLAS:2019bwq,CMS:2017wtu}. We further apply a preselection on the testing set of $N_\mathrm{bjet} > 0$ and $N_\mathrm{jet} \geq 3$ to mimic realistic data analysis requirements.

\section{Results}
\label{sec:results}

\begin{figure}
    \centering
\includegraphics[width=0.45\textwidth]{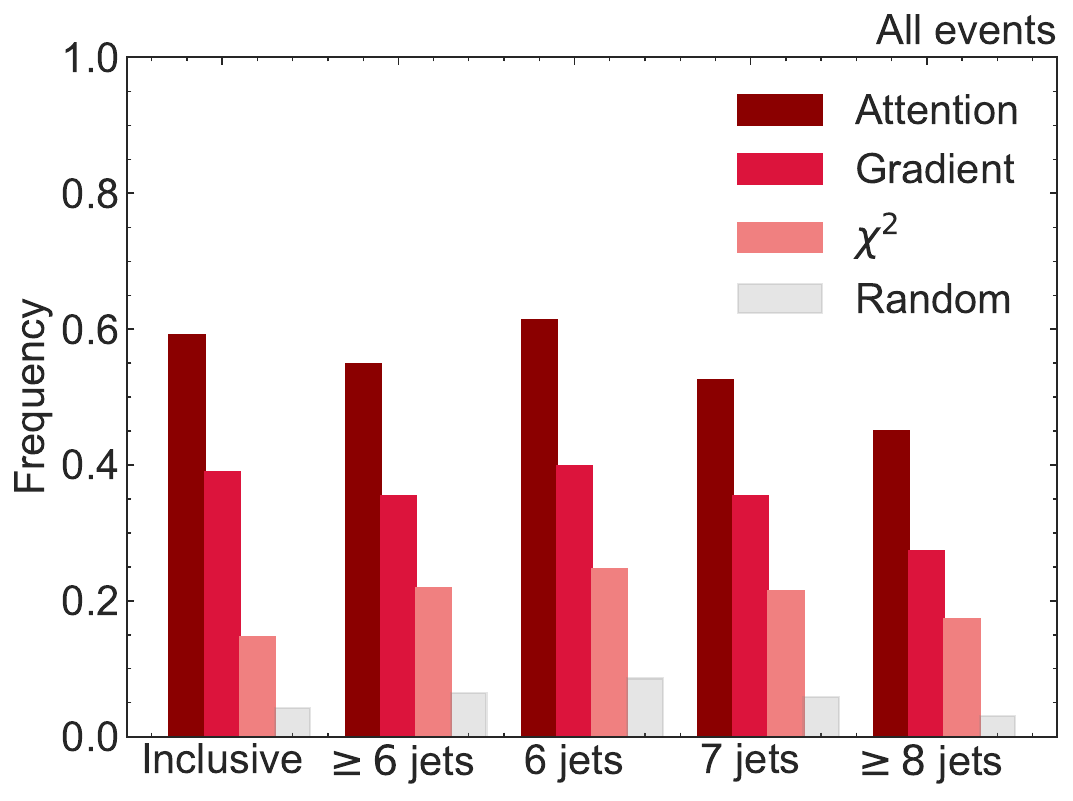}\\
\includegraphics[width=0.45\textwidth]{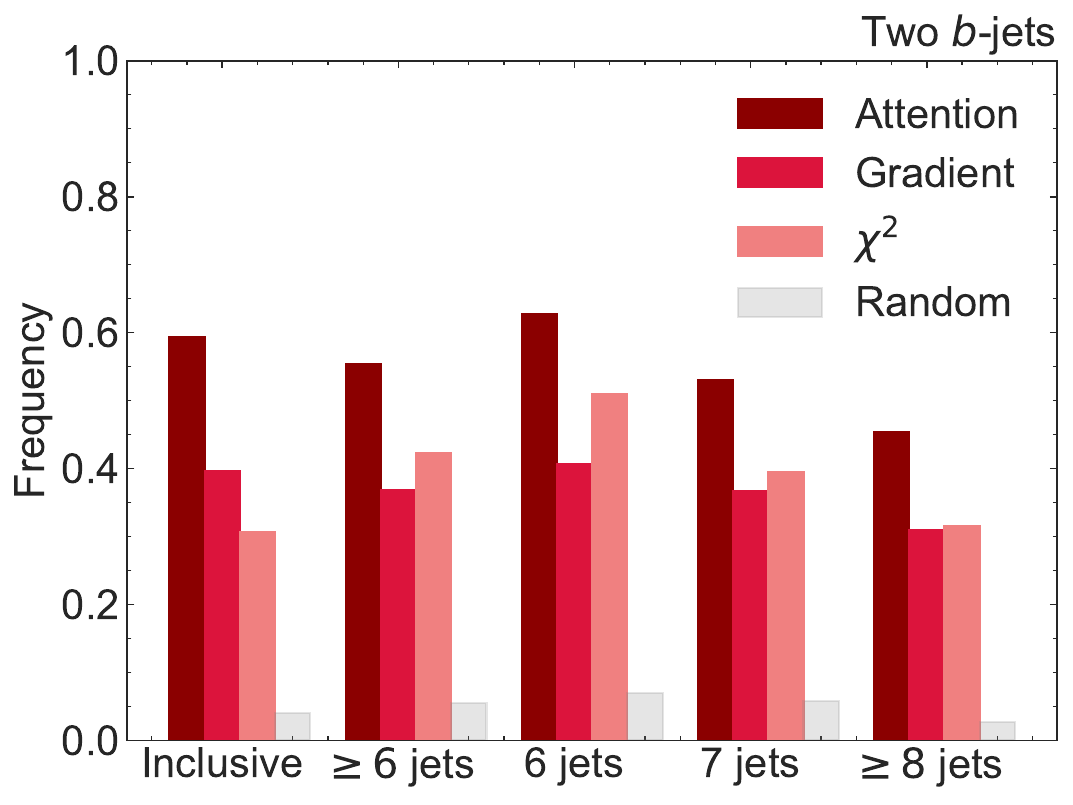}
    \caption{The fraction of truth-matched tops that have exactly the same labels from the truth matching and from the indicated reconstruction method.  Note that these truth labels are not unique, but this is a standard metric.  Top: all events that pass the preselection.  Bottom: only events with at least two $b$-jets. Random corresponds to events with at least six jets and from these, two sets of three are randomly selected.}
    \label{fig:truthmatch}
\end{figure}

First, we consider standard, non-unique metrics for evaluating performance.  In particular, truth-matched top quarks are compared with each reconstruction method to see the fraction of the time that all three jets are the same.  As noted earlier, the truth match labels are not unique, but this is a standard metric for quantifying performance.  Figure~\ref{fig:truthmatch} shows the frequency of an exact match for each method and for different jet multiplicities.  The matching generally is harder the more jets there are in the event because there are more combinations and the truth label fidelity also degrades (see Fig.~\ref{fig:wjets}).

Overall, the attention-based approach outperforms the other two methods across all configurations, often by a large margin (10\% or more).  Inclusively, the gradient-based method outperforms the classical $\chi^2$ assignment, but the two approaches are comparable after requiring two $b$-jets.  Across all events and inclusively across jet multiplicities, the $\chi^2$ approach has a poor matching frequency (about 10\%) in part because it requires two $b$-jets and at least six distinct jets.  In contrast, the attention- and gradient-based methods are still effective when there are fewer jets.  The numbers for the attention-based and $\chi^2$-based approaches are similar to the ones found by \textsc{Spa-Net}~\cite{Fenton:2020woz}, although there are a number of differences in the setup that prohibit a precise comparison.

\begin{figure}
    \centering
\includegraphics[width=0.45\textwidth]{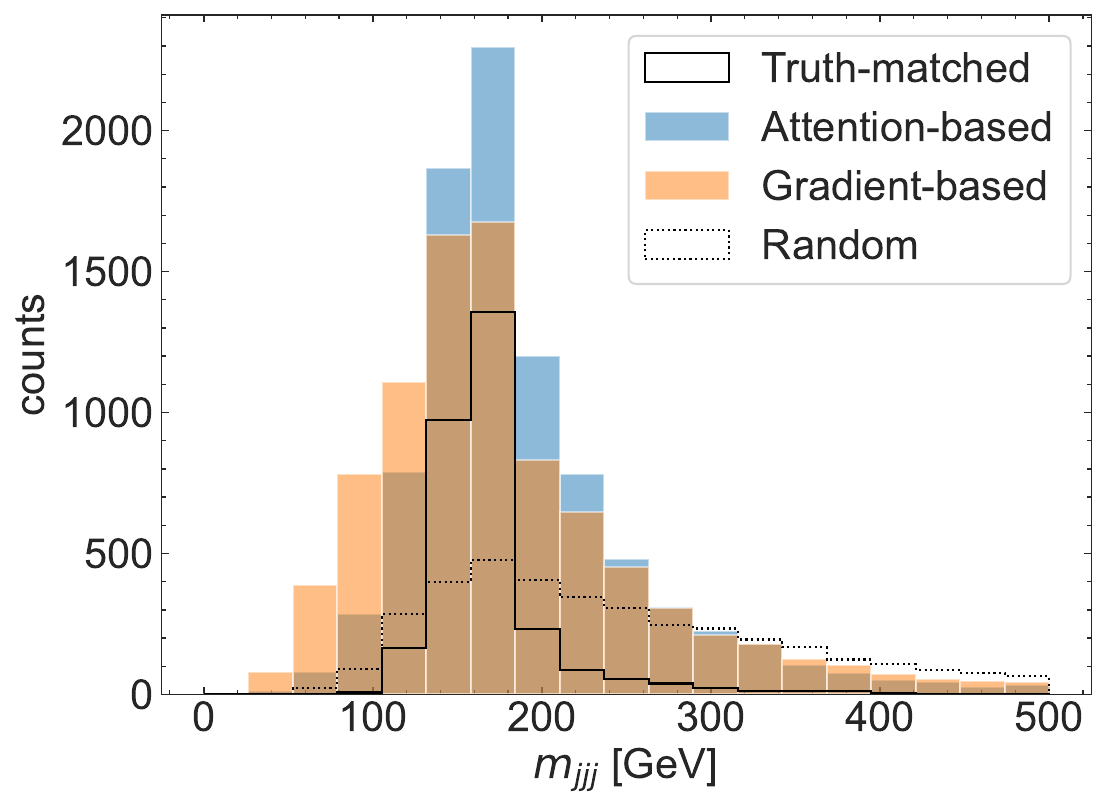}
\includegraphics[width=0.45\textwidth]{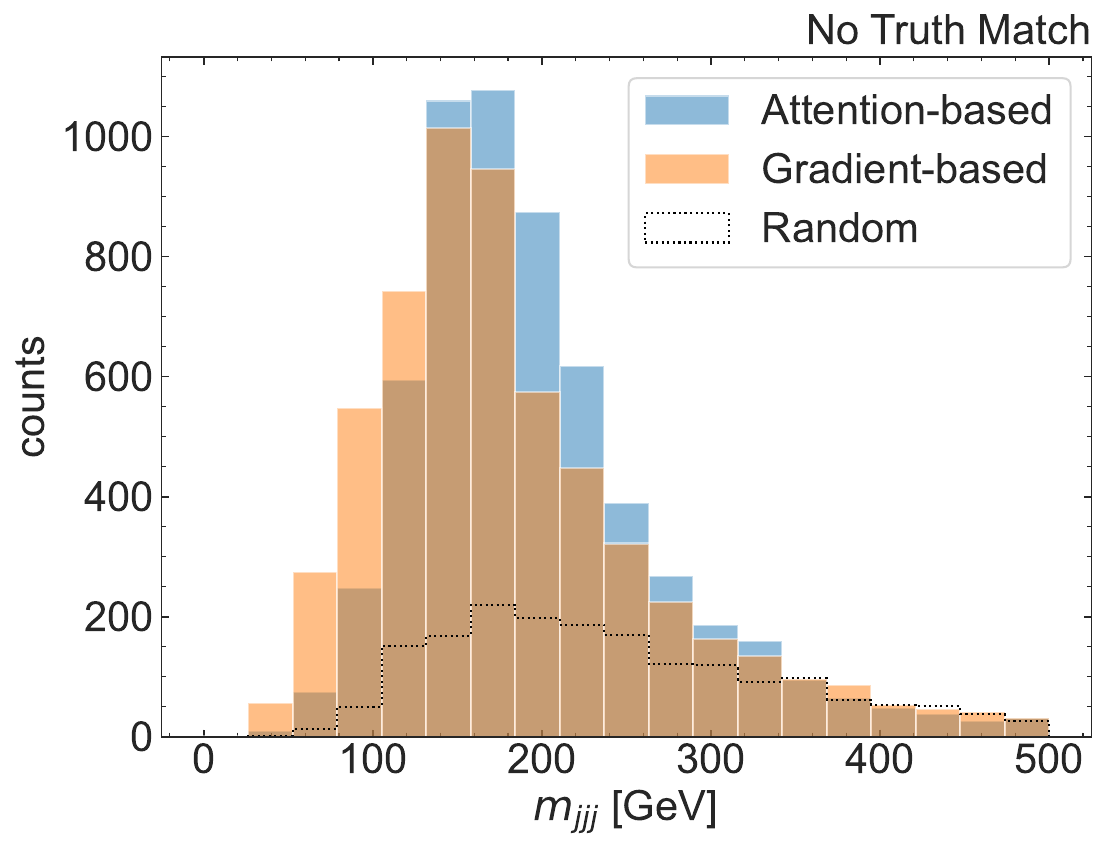}
    \caption{The invariant mass of jets labeled as originating from the same top quark in all events (top) and in events without a truth match (bottom). Random corresponds to events with at least six jets and from these, two sets of three are randomly selected. }
    \label{fig:mjjj}
\end{figure}

The next question is to study events in which there is no truth-match.  Such events are not even part of the training for other ML-based labeling schemes, but our methods are still able to assign parton labels in these cases.  One way to see if the assigned jets in such events are sensible is to examine their trijet invariant mass.  Figure~\ref{fig:mjjj} presents histograms of this map inclusively and for events without a truth match.  There are roughly twice as many entries for the attention- and gradient-based histograms in the top plot of Fig.~\ref{fig:mjjj} because of events where there is no truth match.  All five histograms in the figure look similar, with a peak near the top quark mass of about 175 GeV~\cite{10.1093/ptep/ptaa104}.  The peak sharpest for the truth-matched events and is slightly sharper for the attention-based method than the gradient-based method.  This may be expected from Fig.~\ref{fig:truthmatch}, which indicates that the attention-based approach has a higher fidelity of picking the `correct' jets.

\begin{figure}
    \centering
\includegraphics[width=0.45\textwidth]{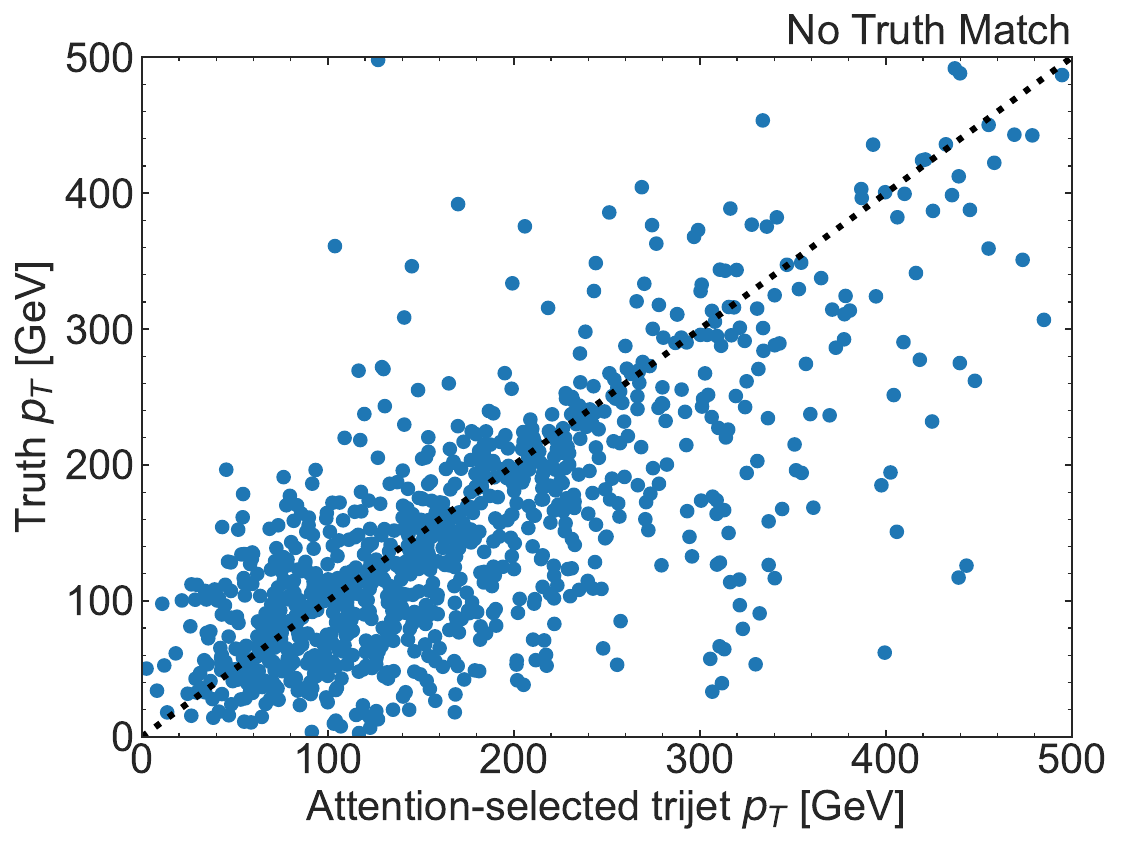}
\includegraphics[width=0.45\textwidth]{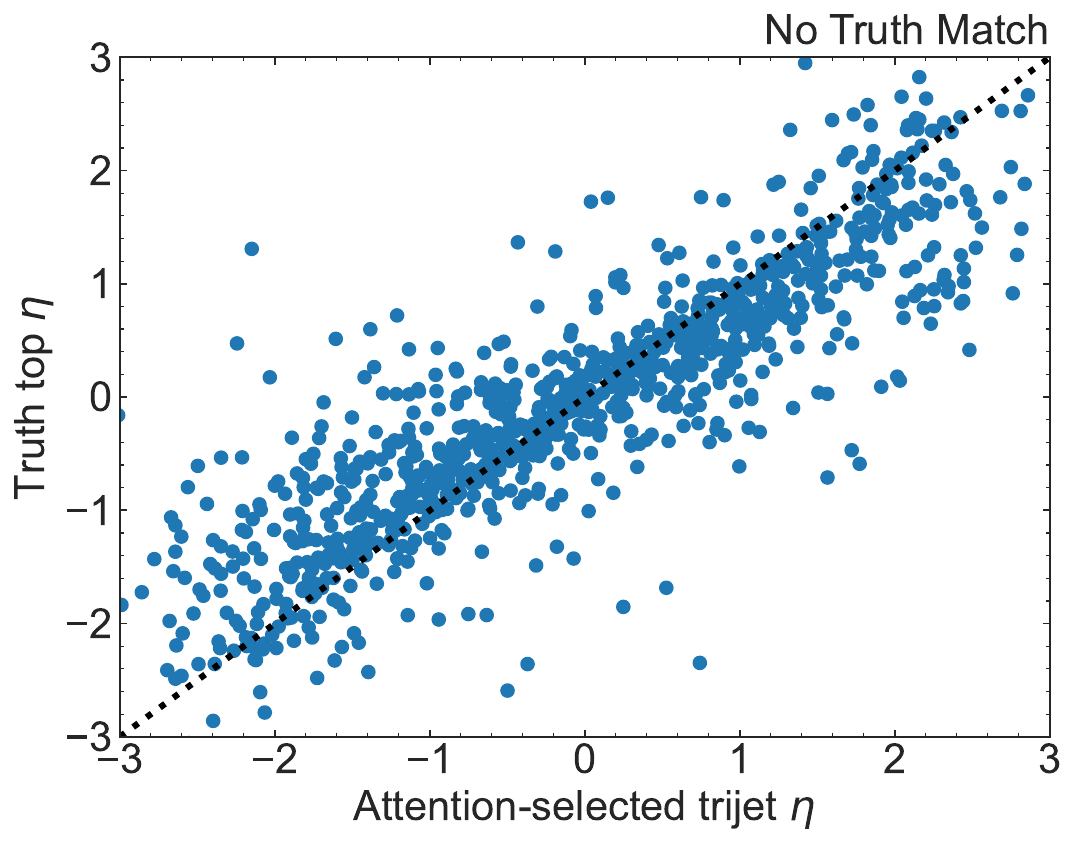}
    \caption{Scatter plots between the true top quark properties ($y$-axis) and the trijet kinematic properties ($x$-axis) for the $p_T$ (top) and $\eta$ (bottom).  None of these events have a truth match.  Versions with random jets are presented in Fig.~\ref{fig:corrrand}.}
    \label{fig:corr}
\end{figure}

Our last investigation is if the trijet kinematic properties in unmatched events are close to the truth top quarks.  One reasonable definition of a `good match' would be that the reconstructed top properties are close to the truth properties, which does not require assigning quark identities to the jets.  Since our methods are derived from a top quark property regressor, we would expect that the trijet properties align well with the truth top quark properties, but it is important to check.  Figure~\ref{fig:corr} provides confirmation for the top quark $p_T$ and $\eta$.

\section{Conclusions and Outlook}
\label{sec:end}

Parton labeling continues to be an important task in collider event reconstruction even though such labels are not unique.  We have proposed a set of tools based on regression methods that are able to assign parton labels without also needing unphysical parton matching for training.  Our approaches are competitive even though they are not trained using trijet information and are much more flexible than other approaches, since we are able to accommodate events with fewer jets than expected from the lowest order decay Feynman diagrams.   While our techniques are compatible with many regression approaches, the CPT model studies here is particularly useful because it is permutation invariant and partially Lorentz covariant.  The corresponding labels inherit some of these properties.

There are a number of possible ways to further improve these approaches, including how to best combine the attention weights or Jacobian elements to assign parton labels.  It may also be possible to combine approaches in the future, where a simpler model can be trained using the label information from a regression model.  

\section*{Software}
The code for this project is built on the one from Ref.~\cite{Qiu:2022xvr}.  Updated software that produces also the gradients and makes the figures in this paper can be found at \url{https://github.com/hep-lbdl/Covariant-Particle-Transformer}.

\section*{Acknowledgments}
BN thanks Chase Shimmin for useful discussions. This work is supported by the U.S.~Department of Energy, Office of Science under contract DE-AC02-05CH11231. H.W.'s work is partly supported by the U.S. National Science Foundation under the Award No. 2046280.


\bibliography{references,HEPML}
\newpage
\appendix

\begin{figure}
    \centering
\includegraphics[width=0.45\textwidth]{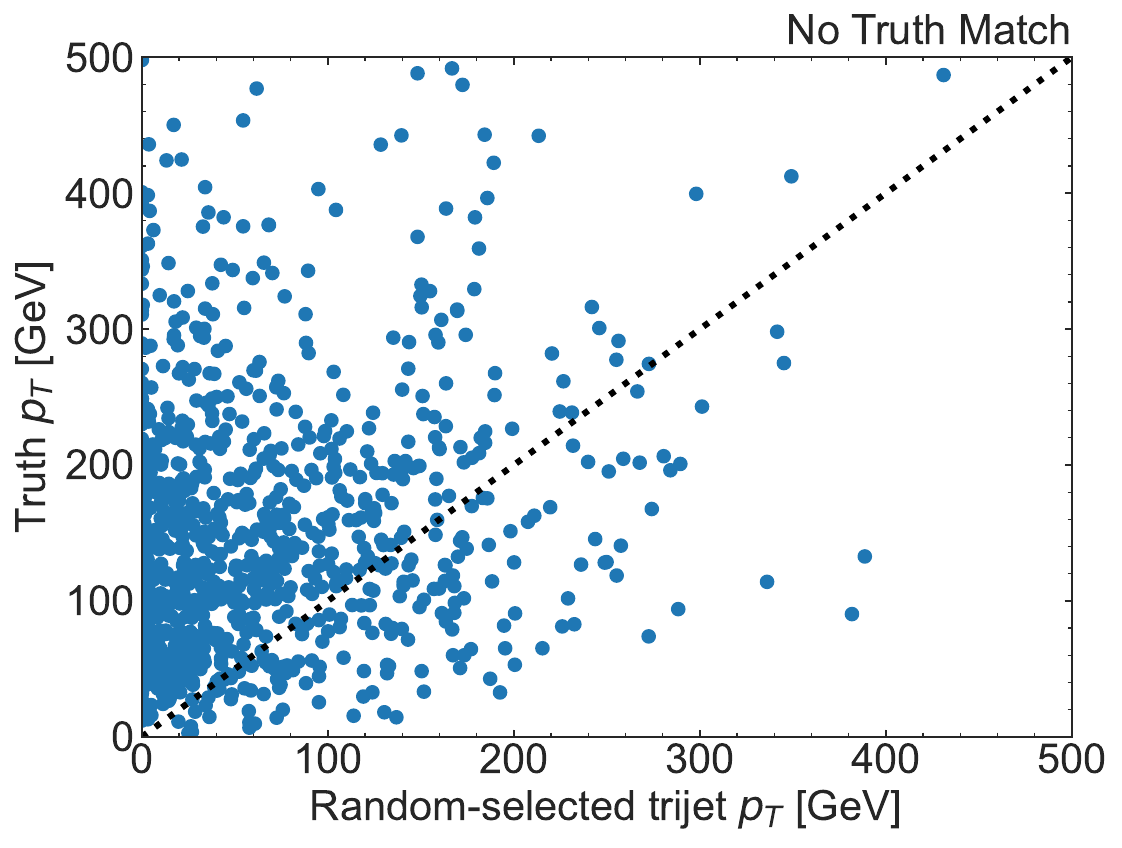}
\includegraphics[width=0.45\textwidth]{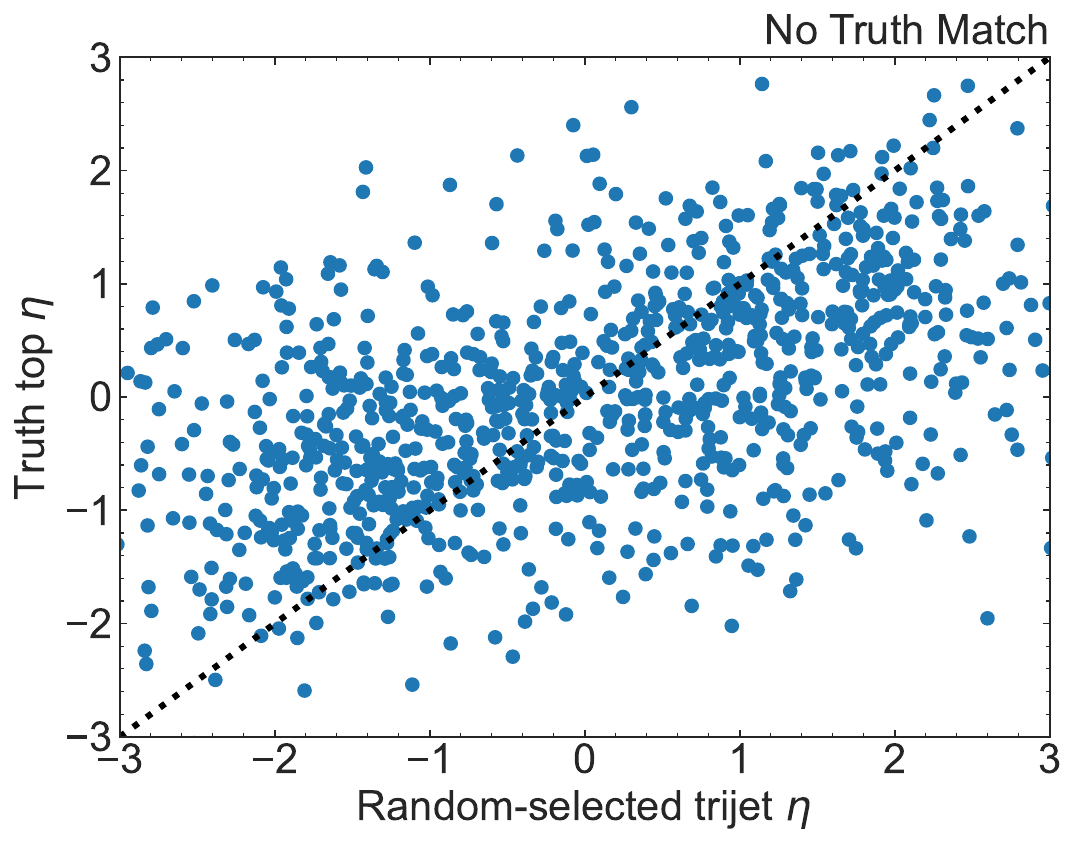}
    \caption{Scatter plots between the true top quark properties ($y$-axis) and the trijet kinematic properties ($x$-axis) for the $p_T$ (top) and $\eta$ (bottom) using three randomly selected jets for each top candidate.  None of these events have a truth match.}
    \label{fig:corrrand}
\end{figure}

\end{document}